\documentclass[aip,jcp,amsmath,amssymb,reprint,nofootinbib,floatfix,longbibliography]{revtex4-2}

\usepackage[utf8]{inputenc}
\usepackage[T1]{fontenc}
\usepackage{lmodern}
\usepackage{mathtools}
\usepackage{bm}
\usepackage{graphicx}
\usepackage{xcolor}
\usepackage{tikz}
\usepackage{pgfplots}
\pgfplotsset{compat=1.17}
\usetikzlibrary{arrows.meta,calc,patterns}
\usepackage[colorlinks=true,linkcolor=blue!60!black,citecolor=blue!60!black,urlcolor=blue!60!black]{hyperref}
\usepackage{orcidlink}
\usepackage{placeins}
\usepackage{float}

\makeatletter
\let\@orcidqueue\@empty
\newcommand{\orcid}[1]{\g@addto@macro\@orcidqueue{\@oitem{#1}}}
\def\popfirst@orcid{%
  \ifx\@orcidqueue\@empty\else
    \expandafter\popfirst@orcid@aux\@orcidqueue\@orcidSTOP
  \fi
}
\def\popfirst@orcid@aux\@oitem#1#2\@orcidSTOP{%
  \,\mbox{\orcidlink{#1}}%
  \gdef\@orcidqueue{#2}%
}
\def\doauthor#1#2#3{%
  \ignorespaces#1\unskip\@listcomma
  \begingroup
   #3%
  \@if@empty{#2}{\endgroup{}{}}{\endgroup{\comma@space}{}\frontmatter@footnote{#2}}%
  \popfirst@orcid
  \space \@listand
}%
\long\def\frontmatter@footnotetext#1{%
  \par
  \begingroup
    \set@footnotefont
    \frontmatter@makefntext{\ignorespaces#1}%
  \endgroup
  \par
}%
\makeatother

\newcommand{\phigold}{\varphi}                  
\newcommand{\Jcost}{J}                          
\newcommand{\dlay}{d}                           
\newcommand{\Lrod}{\ell}                        
\newcommand{\Drod}{D}                           
\newcommand{\nhat}{\hat{\mathbf{n}}}            

\newcommand{\Rsph}{R_{\mathrm{s}}}              
\newcommand{\eyhat}{\hat{\mathbf{e}}_\theta}    
\newcommand{\ephihat}{\hat{\mathbf{e}}_\phi}    
\newcommand{\tilt}{\omega}                      
\newcommand{\legball}[1]{\protect\tikz[baseline=-0.55ex]{\protect\shade[ball color=#1] (0,0) circle (0.62ex);}}

\newcommand{\rNEW}[1]{#1}

\newcounter{rissue}

\begin{document}

\raggedbottom

\title{Curvature-induced smectic-C order of tangentially anchored
       hard spherocylinders on a sphere
       \rNEW{with a rigidly locked director field}}

\author{Jonathan Washburn}
\orcid{0009-0001-8868-7497}
\affiliation{Recognition Physics Institute, Austin, TX, USA}

\author{Hartmut L\"owen}
\orcid{0000-0001-5376-8062}
\affiliation{Institut f\"ur Theoretische Physik II: Weiche Materie,
  \mbox{Heinrich-Heine-Universit\"at D\"usseldorf, Germany}}

\author{Elshad Allahyarov}
\orcid{0000-0001-7212-4713}
\email[Author to whom correspondence should be addressed: ]{elshad.allakhyarov@case.edu}
\affiliation{Recognition Physics Institute, Austin, TX, USA}
\affiliation{Institut f\"ur Theoretische Physik II: Weiche Materie,
  \mbox{Heinrich-Heine-Universit\"at D\"usseldorf, Germany}}
\affiliation{Department of Physics, Case Western Reserve University,
  \mbox{Cleveland, OH 44106, USA}}
\affiliation{Theoretical Department, Joint Institute for High Temperatures,
  RAS, \mbox{Moscow 125412, Russia}}

\begin{abstract}
We study the strict locked-orientation limit of hard spherocylinders on a
sphere, in which the rod axes are rigidly locked to a prescribed
tangential director field and cannot reorient. Because the bulk hard-rod
phase diagram contains no smectic-C phase, any coherent tilt isolates a
geometric curvature mechanism rather than a finite-stiffness
equilibrium effect. A ratio-symmetric recognition cost fixes the layer
spacing at the bulk close-contact value and yields a hierarchy of
geometric statements: the lower edge of the smectic-area window at
$45^\circ$ follows from reciprocal symmetry; the upper edge at
$58.3^\circ$ is a falsifiable channel-saturation hypothesis; the
smectic-A to smectic-C boundary is a closed-form prediction; and
the rod tilt angle is set by the rod-to-radius ratio, modulated by a chirality
envelope peaking near $24^\circ$. Locked-orientation Monte Carlo across
fifteen geometries confirms these predictions with no fitted elastic
constants: the smectic area peaks at $55^\circ$, and a coherent smectic-C
window is detected.
\end{abstract}

\maketitle

\section{Introduction}\vspace{-0.4cm}

Hard spherocylinders are the minimal model of lyotropic liquid
crystals: entropy alone drives the isotropic--nematic transition
\citep{Onsager1949}, and the bulk phase diagram
\citep{Frenkel1988,Bolhuis1997}---isotropic, nematic,
smectic A (Sm-A), columnar, and crystal as a function of aspect ratio
$\Lrod/\Drod$ and packing fraction $\eta$---is reproduced
experimentally by rod-like \emph{fd}-virus suspensions
\citep{Dogic1997}. In particular, that bulk phase diagram contains
\emph{no smectic C (Sm-C) phase}: hard-core repulsion alone does not tilt
the layers in flat space
\citep{Bolhuis1997,deGennesProst,Lagerwall2012}. Any Sm-C
order found in a hard-rod fluid must therefore be induced by an
external geometric frustration.

A rod fluid tangentially anchored on an $S^{2}$ sphere has no remaining
continuous translational symmetry, and the host is topologically
non-trivial: by the Poincar\'e--Hopf theorem
($\chi(S^{2})\!=\!+2$), any tangential director field must carry
singular polar regions of total winding $+2$, so smectic layers
perpendicular to the director cannot close smoothly around the
sphere. Global smectic order on $S^{2}$ is therefore obstructed, and
smectic order is expected to be restricted to patches
around the polar defects \citep{Smallenburg2016,Trukhina2008}. Particle-resolved studies of
this system with \emph{externally locked} orientations exist at the
two extreme tilts only---rods along the meridians or along the
latitudes \citep{Allahyarov2017,Allahyarov2018}---and the
contemporary soft-rod molecular dynamics
\citep{Mandal2026} fixes the meridional orientation
throughout.

Experimentally, related curvature-frustrated ordering has been
studied most thoroughly in liquid-crystal shells produced by
microfluidics. Ref.\onlinecite{FernandezNieves2007} first imaged the
defect structures of nematic double-emulsion shells, finding
coexisting configurations set by the finite, inhomogeneous shell
thickness; Ref.\onlinecite{LopezLeon2011}
showed that the number and arrangement of the topologically required
defects can be tuned between the four-, three-, and two-defect states,
and Ref.\onlinecite{LopezLeonBates2012} that defect pairs coalesce reproducibly
when the surface anchoring is switched. In
the smectic phase the one-dimensional layering conflicts with the
curvature: Ref.\onlinecite{Liang2011} tracked the nematic--smectic
transition in colloidal shells and the buckling instabilities it
triggers, Ref.\onlinecite{LopezLeon2012} resolved the
resulting curvature walls and crescent domains,
and Ref.\onlinecite{Liang2012} obtained tunable focal-conic arrays in
hybrid-aligned smectic shells. Most directly,
Ref.\onlinecite{Sharma2024} showed that smectic-C shells relieve this
frustration through a director tilt, motivating the tilted-layer
mechanism studied here. Closely analogous physics
appears in colloidal membranes of rod-like \emph{fd} virus, which
self-assemble into one-rod-length-thick chiral smectic monolayers
\citep{Barry2010}: their edges and internal rafts develop a
spontaneous rod tilt of definite handedness
\citep{Zakhary2014,Sharma2014}, and even achiral chromonic rods break
chiral symmetry purely through curved confinement \citep{Tortora2011}.

On the modelling side, ordering on curved surfaces has been studied
extensively by theory and simulation. Ref.\onlinecite{Shin2008}
performed Monte Carlo simulations of hard spherocylinders confined to
the tangent plane of a sphere---the closest precedent to the present
system---and found the four $+1/2$ disclinations arranged on a great
circle in the bend-dominated limit. Ref.\onlinecite{BatesJCP2008} simulated
nematic ordering and defect arrangements on a sphere by Monte Carlo,
and Ref.\onlinecite{Bates2010} extended this
to nematic coatings on uniaxial and biaxial colloids;
Ref.\onlinecite{Dhakal2012} showed how the
defect geometry flows from tetrahedral to great-circle arrangements as
temperature, density, and rod shape tune the effective elastic
anisotropy. On the continuum side, Ref.\onlinecite{Vitelli2006}
derived the equilibrium textures of nematic shells as a
function of thickness; Refs.\onlinecite{Kralj2011} and~\onlinecite{Koning2016}
analysed how curvature controls the defect
valence and stabilises three-defect shells;
and Refs.\onlinecite{Napoli2012,Napoli2012b} established that the extrinsic curvature of the
shell couples to the director through the Frank constants. The broader interplay of order,
curvature, and topological defects in two dimensions is reviewed in
Ref.\onlinecite{Bowick2009}.

The intermediate-tilt window, where the locked director
itself is chiral on the sphere and the layer geometry must
accommodate both meridional and latitudinal winding, is the
subject of this Communication.
Prior particle-resolved work on this minimal lyotropic system has
reported nematic and smectic-A order on the sphere but not a
curvature-induced smectic-C transition; the intermediate-tilt regime in
which that transition appears has not previously been mapped.

We show that the curvature of a closed spherical host is exactly such a
geometric frustration, we derive the parameter-free geometric scales of the
resulting curvature-induced Sm-C phase, and we confirm them by
particle-resolved Monte Carlo (MC) simulation. Curvature here is
the sole driver: the tilt is induced with no chirality or amphiphilicity
input, in contrast to the chiral-smectic shells and rod-like
colloidal membranes in which the layer tilt is set by molecular
chirality.

\section{Theory}\vspace{-0.4cm}

The system is a fluid of monodisperse hard spherocylinders (cylinder
length $\Lrod$, diameter $\Drod$) whose centres lie on a sphere of radius
$\Rsph$ and whose axes are rigidly locked to the tangential director
field
\begin{equation}
 \nhat_{\tilt}(\theta,\phi)=\cos\tilt\,\eyhat+\sin\tilt\,\ephihat,
 \label{eq:director}
\end{equation}
with $\eyhat$ the meridian direction, $\ephihat$ the latitude direction,
and $\tilt$ the imposed director tilt. Thus the control variables relevant
to the present Communication are the locked tilt $\tilt$ and the curvature
ratio $(\Lrod+\Drod)/\Rsph$.

We use the ratio-symmetric recognition cost of the Recognition Science (RS) 
framework~\citep{WashburnRSnotes,WashburnLedger},
\begin{equation}
 \Jcost(x)=\frac{1}{2}\left(x+\frac{1}{x}\right)-1,
 \qquad x>0,
 \label{eq:Jdef}
\end{equation}
as a phenomenological cost density for comparing layer mismatches on the
curved host. The use of Eq.\eqref{eq:Jdef} in smectic layer energetics is
a modelling assumption, not a derivation from hard-rod free energy. Its
role here is to generate parameter-free geometric tests for the
locked-orientation simulations; the derivations and assumptions are given
in the Supplementary Material (SM). As a consistency check, the matched scale of
Eq.\eqref{eq:Jdef} gives the standard hard-rod smectic spacing
$d^*=\Lrod+\Drod$.

Three geometric statements (P1--P3) stemming from the RS framework are used below. First (P1), the
smectic-area window is obtained from the competition of meridional and
latitudinal layer-breaking channels (see the SM for details),
\begin{equation}
 \mathcal{C}_{\rm sm}(\tilt)=
 w_{\rm A}\,\Jcost(\phigold \cdot y_{\rm A})+
 w_{\rm B}\,\Jcost(\phigold \cdot  y_{\rm B}),
 \label{eq:Csm}
\end{equation}
where $w_{\rm A}=\sin^2\tilt$, $w_{\rm B}=\cos^2\tilt$,
$y_{\rm A}\propto1/\sin\tilt$, and
$y_{\rm B}\propto1/\cos\tilt$. The lower edge follows from reciprocal
symmetry of $J$ at equal channel weight, whereas the upper edge invokes the
channel-saturation hypothesis $w_{\rm B}/w_{\rm A}=1/\phigold^2$:
\begin{equation}
 \tilt_{\rm sm}^{\ast}\in
 \bigl[\pi/4,\arctan\phigold\bigr]
 \approx[45^\circ,58.3^\circ].
 \label{eq:SP4}
\end{equation}

Second (P2), a tilted Sm-A layer geometry spirals around latitude circles,
whereas a Sm-C layer geometry closes on latitude circles at the cost of a
rod projection mismatch. Equating these two costs with the reciprocal
symmetry of Eq.\eqref{eq:Jdef} gives the Sm-A to Sm-C boundary
\begin{equation}
 R^{\ast}_{\rm SmC}(\tilt)=
 \frac{\Lrod+\Drod}{2\pi\sin\tilt\,(\sec\tilt-1)}.
 \label{eq:SP5}
\end{equation}
Its monotonic decrease with $\tilt$ places the
coherent Sm-C region at low tilt (below $\pi/4$).

Third (P3), once a Sm-C layer forms, the leading tilt angle of the rods
is set by the rod-to-radius ratio,
\begin{equation}
 |\alpha^{\ast}|(\Rsph,\Lrod)=
 \arctan\left(\frac{\Lrod+\Drod}{\Rsph}\right),
 \label{eq:SP6}
\end{equation}
while the locked tilt sets whether that tilt points the same way around
the whole sphere, that is, whether it has a single handedness. The
average tilt $\langle\alpha\rangle$, taken with its sign so that it is
nonzero only when one handedness wins, is zero at $\tilt=0$ and
$\tilt=\pi/2$ and is largest at small imposed tilt rather than at the
midpoint $\pi/4$. We describe how it varies with $\tilt$ by
\begin{equation}
 \langle\alpha\rangle(\tilt)\propto
 \pm |\alpha^{\ast}|\sin(2\tilt)\cos^{\nu}\!\tilt,
 \qquad \nu=\phigold^3,
 \label{eq:shift}
\end{equation}
whose maximum lies at
\begin{equation}
 \tilt_{\rm pk}=\arctan\frac{1}{\phigold\sqrt{2}}
 \approx23.6^\circ.
 \label{eq:peak}
\end{equation}
The exponent in Eq.\eqref{eq:shift} is calibrated to the observed
low-tilt maximum; it is therefore $\phigold$-consistent, not a separate
parameter-free prediction. Evaluating Eq.\eqref{eq:SP5} near this peak
gives the coherent-block scale
\begin{equation}
 R^{\ast}_{\rm coh}\equiv R^{\ast}_{\rm SmC}(\tilt_{\rm pk})
 \approx5(\Lrod+\Drod),
 \label{eq:Rcoh}
\end{equation}
which is used only as a geometric scale for the radial edge.

\section{Simulation Methods}\vspace{-0.4cm}

\begin{figure}[H]
\centering
\includegraphics[width=0.95\linewidth]{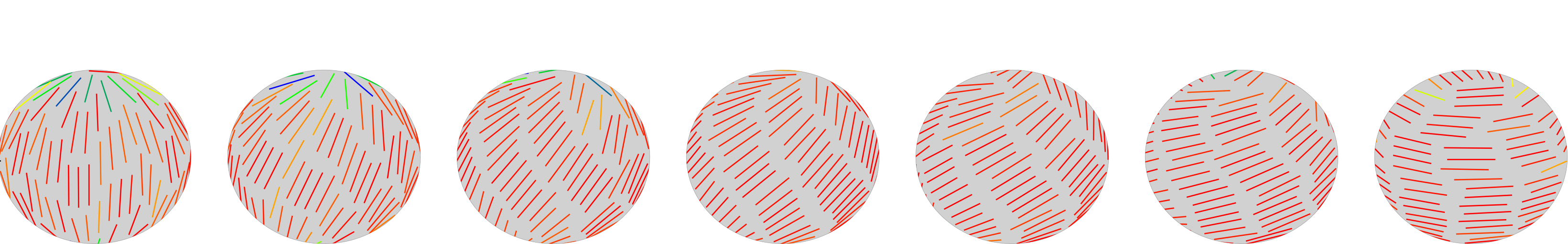}\\[1pt]
\includegraphics[width=0.95\linewidth]{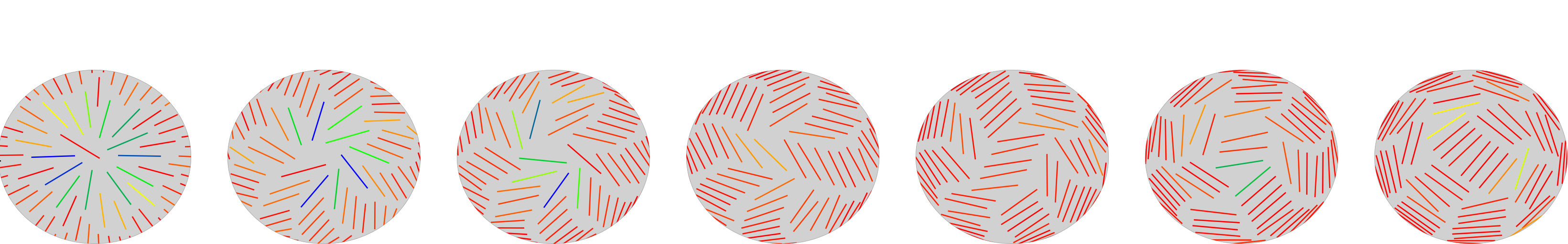}\\[1pt]
\includegraphics[width=0.95\linewidth]{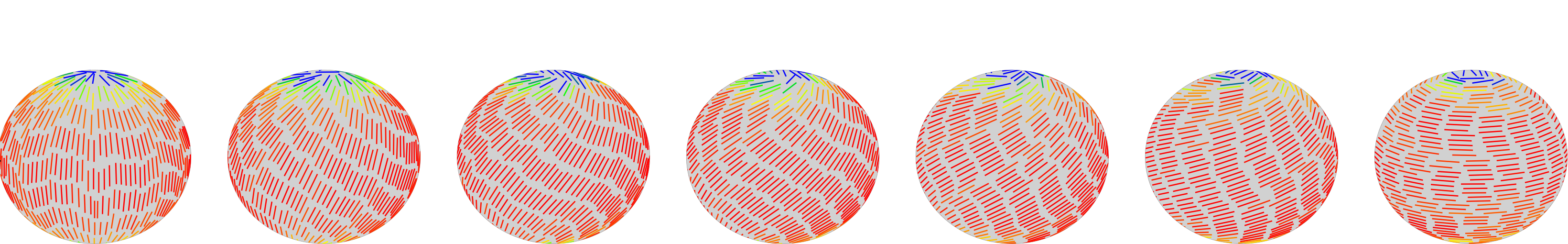}\\[1pt]
\includegraphics[width=0.95\linewidth]{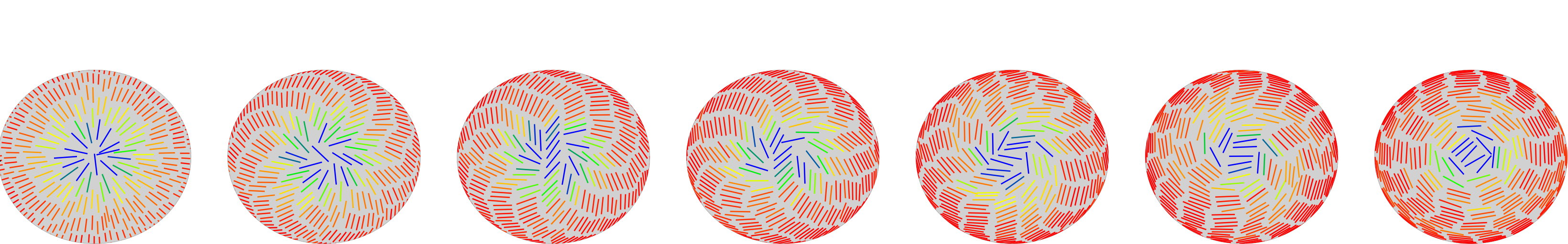}
\caption{Simulation snapshots at the locked tilts $\tilt=0$,
$22.5^{\circ}$, $35^{\circ}$, $45^{\circ}$, $55^{\circ}$,
$67.5^{\circ}$, $90^{\circ}$ (left to right), for two hosts.
\emph{Top two rows:} the smallest host $\Rsph=10\Drod$,
$\Lrod=5\Drod$,  side view (first row) and
top-down (north-pole-down) view (second row).
\emph{Bottom two rows:} the same but for the coherent-Sm-C host $\Rsph=20\Drod$,
$\Lrod=5\Drod$. 
 Rods are coloured
 by their local nematic order parameter.
}
\label{fig:snapR10}
\end{figure}

We sample the system by strict
locked-orientation NVT MC, the exact
$K_{3}/K_{1}\!\to\!\infty$ realisation of the bend-stiff limit,
with $K_{1}$ and $K_{3}$ the Frank splay and bend elastic
constants of the director field, so that the prescribed tilt in
Eq.\eqref{eq:director} is held rigid against all orientational
fluctuations; every accepted configuration is provably overlap-free by
direct Vega--Lago re-evaluation~\citep{VegaLago1994}. The study spans fifteen
$(\Rsph,\Lrod)$ geometries [all combinations of
$\Rsph\in\{10,20,30,40\}\Drod$ and $\Lrod\in\{3,5,8,10\}\Drod$
except the frustrated corner $(\Rsph,\Lrod)=(10,10)\Drod$, where
the rod length equals the host radius ($\Lrod/\Rsph=1$, far above
the geometric smectic-formation bound $\Lrod/\Rsph\sim 1/3$)],
packing
fraction $\eta=0.75$ (at which smectic order reliably
forms~\citep{Allahyarov2017,Allahyarov2018}), and up to
twelve tilt angles per geometry.  All simulation details
are collected in the SM.
Figure~\ref{fig:snapR10} shows some representative
snapshots, both side-views and top-views
where the locked chiral
director field and the banded layering at intermediate tilts are
visible. 
The smectic observable is local, as the topology demands.

For each rod $i$ with neighbour set $\mathcal{N}_{i}$ (centre
distance below $r_{\rm cut}$) we define
\begin{equation}
 \Psi_{i}(\dlay) = \frac{1}{|\mathcal{N}_{i}|}
 \sum_{j\in\mathcal{N}_{i}}
 \exp\!\Bigl[2\pi\,\mathrm{i}\,
 \frac{(\mathbf{r}_{j}-\mathbf{r}_{i})\cdot\hat{m}_{i}}{\dlay}\Bigr],
 \label{eq:psi}
\end{equation}
where the trial layer normal $\hat{m}_{i}$ is tilted by an angle
$\alpha$ from the locked director in the local tangent plane, and
$|\Psi_{i}|$ is jointly maximised over the trial spacing $\dlay$
and the layer tilt $\alpha$, giving the per-rod optima
$(|\Psi_{i}^{\ast}|, d_{i}^{\ast}, \alpha_{i}^{\ast})$. Setting
$\alpha\!=\!0$ recovers the standard Sm-A order parameter. The neighbour
distances are ambient Euclidean chord distances on the particle centres,
and the neighbour cutoffs, trial-spacing ranges, and the numbers of trial
spacings and tilts used in the joint optimisation are listed in the
SM.
Three aggregates summarise a panel: the smectic area fraction
$\chi_{\rm sm}$ (fraction of rods with $|\Psi_{i}^{\ast}|>0.5$, a
threshold midway between the disordered and fully layered limits),
the median local spacing $\tilde d^{\ast}$, and the chirality
index $\chi\equiv|\langle\alpha\rangle|/\sigma_{\alpha}$, where
$\langle\alpha\rangle$ and $\sigma_{\alpha}$ are the panel mean
and standard deviation of the per-rod layer tilts
$\alpha_{i}^{\ast}$. It
discriminates a globally coherent Sm-C with a single
handedness ($\chi\ge 1.5$) from a Sm-A or multi-domain
configuration in which the signed mean is comparable to the
per-rod scatter. The joint $(d,\alpha)$ optimisation generates
per-rod tilt noise of order $\sigma_{0}\approx8^{\circ}$ on a true Sm-A
configuration, so a nonzero $\langle|\alpha|\rangle$ alone does
not establish Sm-C order, whereas the criterion $\chi\ge1.5$ does.
On a disordered or multi-domain panel the signed mean is small
($\langle\alpha\rangle\!\approx\!0$, hence $\chi\!\approx\!0$), whereas a
single-handed tilt gives $\chi\gg1$; the cut $\chi\ge1.5$ sits between
these regimes.

\section{Results}\vspace{-0.4cm}

 \rNEW{The MC data set supports the RS-predicted lower-edge/window, boundary,
and short-rod tilt-angle statements at their stated status}. We discuss
 the evidence figure by figure.

The layer-spacing fixed point $d^{\ast}$, which is the standard close-contact layer period of
bulk hard-rod smectics \citep{Frenkel1988,Bolhuis1997},
is tested across all strict-MC
panels. The median local spacing
$\tilde d^{\ast}$ lies
within $\pm 10\%$ of $\Lrod+\Drod$ at every curvature, density,
and tilt, for which the dimensionless cost
$\Jcost(\tilde d^{\ast}/(\Lrod+\Drod)) < 6\times10^{-3}$ on every panel.
This
confirms that the cost functional reproduces known
packing. 

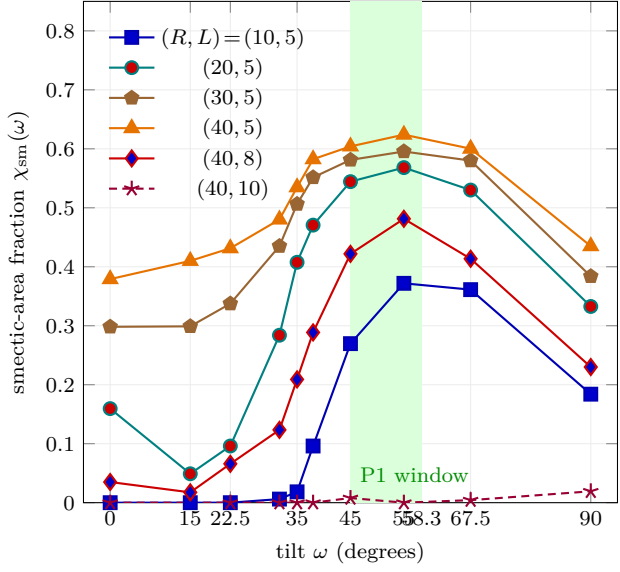
\begin{figure}[H]
\centering
\begin{tikzpicture}
\begin{axis}[
    width=\columnwidth, height=0.95\columnwidth,
    xlabel={tilt $\tilt$ (degrees)},
    ylabel={smectic-area fraction $\chi_{\rm sm}(\tilt)$},
    xlabel style={font=\footnotesize},
    ylabel style={font=\footnotesize},
    xmin=-5, xmax=95,
    ymin=0, ymax=0.85,
    xtick={0,15,22.5,35,45,55,58.3,67.5,90},
    unbounded coords=discard,
    xticklabel style={font=\footnotesize},
    ytick={0,0.1,0.2,0.3,0.4,0.5,0.6,0.7,0.8},
    yticklabel style={font=\footnotesize},
    grid=both, grid style={gray!15, very thin},
    legend pos=north west,
    legend style={font=\footnotesize, draw=none, fill=none},
    title style={font=\footnotesize},
  ]
  \fill[green!14] (axis cs:45,0) rectangle (axis cs:58.3,0.85);
  \node[anchor=south, font=\footnotesize, green!55!black]
    at (axis cs:57,0.02) {P1 window};

  \addplot+[thick, blue!70!black,    mark=square*,   mark size=2.4pt]
    coordinates {(0,0) (15,0.000) (22.5,0) (31.70,0.0061) (35,0.0184)
      (38,0.0960761) (45,0.26967895) (55,0.37193817) (67.5,0.36123662)
      (90,0.184)};
  \addlegendentry{$(R,L)\!=\!(10,5)$}
  \addplot+[thick, teal,              mark=*,         mark size=2.4pt]
    coordinates {(0,0.1595) (15,0.049) (22.5,0.0960761) (31.70,0.28394768)
      (35,0.40760999) (38,0.4706302) (45,0.54435196) (55,0.56813317)
      (67.5,0.53008323) (90,0.33269917)};
  \addlegendentry{$(20,5)$}
  \addplot+[thick, brown!80!black,   mark=pentagon*, mark size=2.6pt]
    coordinates {(0,0.29821641) (15,0.299) (22.5,0.33745541) (31.70,0.43495838)
      (35,0.50630202) (38,0.55148633) (45,0.58121284) (55,0.59548157)
      (67.5,0.58002378) (90,0.38382878)};
  \addlegendentry{$(30,5)$}
  \addplot+[thick, orange!90!black,  mark=triangle*, mark size=3pt]
    coordinates {(0,0.37907253) (15,0.41) (22.5,0.4313912) (31.70,0.48014269)
      (35,0.53483948) (38,0.5824019) (45,0.6041) (55,0.62401902)
      (67.5,0.60023781) (90,0.43495838)};
  \addlegendentry{$(40,5)$}
  \addplot+[thick, red!80!black,     mark=diamond*,  mark size=3pt]
    coordinates {(0,0.035) (15,0.0175) (22.5,0.0659) (31.70,0.12342449)
      (35,0.20903686) (38,0.28870392) (45,0.42187872) (55,0.48133175)
      (67.5,0.41355529) (90,0.23)};
  \addlegendentry{$(40,8)$}
  \addplot+[thick, purple!80!black,  mark=star,      mark size=3pt]
    coordinates {(0,0) (22.5,0) (31.70,0) (35,0) (38,0) (45,0.0079) (55,0)
      (67.5,0.0043) (90,0.0193)};
  \addlegendentry{$(40,10)$}
\end{axis}
\end{tikzpicture}
\caption{Smectic-area fraction $\chi_{\rm sm}(\tilt)$
across six $(\Rsph,\Lrod)$ geometries (fraction of
rods with local smectic amplitude $|\Psi_{i}^{\ast}|>0.5$).
Shaded band: the \rNEW{RS} smectic-area (P1) window
$[45^{\circ},58.3^{\circ}]$ [Eq.\eqref{eq:SP4}].
\rNEW{The lower edge follows from reciprocal symmetry; the upper edge is
the channel-saturation hypothesis and is not resolved separately by the
present tilt grid.}}
\label{fig:chisummary}
\end{figure}

Figure~\ref{fig:chisummary} shows the smectic-area fraction
$\chi_{\rm sm}(\tilt)$ across the
$\Lrod=5\Drod$ radius series [$(10,5)$, $(20,5)$, $(30,5)$, $(40,5)$] and
the two longer-rod cohorts $(40,8)$ and $(40,10)$. Two
features establish P1. First, every 
cohort peaks at the \emph{same} tilt $\tilt=55^{\circ}$, which is   
inside the $[45^{\circ},58.3^{\circ}]$ window
of Eq.\eqref{eq:SP4}, a
parameter-free target fixed entirely by the recognition cost.
Note that Eq.\eqref{eq:Csm} predicts the band, not a single peak angle. 
Second, the $(40,10)$
curve is flat at $\chi_{\rm sm}\le 2\%$, meaning that the smectic channel
is entirely suppressed on the longest rods. This is consistent with the finding of Ref.\onlinecite{Mandal2026}:
the long-rod aspect ratio $\Lrod=10\Drod$ exceeds the critical
aspect ratio $\Lrod/\Drod \in (8,9)$, above which
spherical confinement suppresses smectic ordering.
Hence neither the spiralling Sm-A nor the
tilted Sm-C orders, and the smectic channel
closes. 

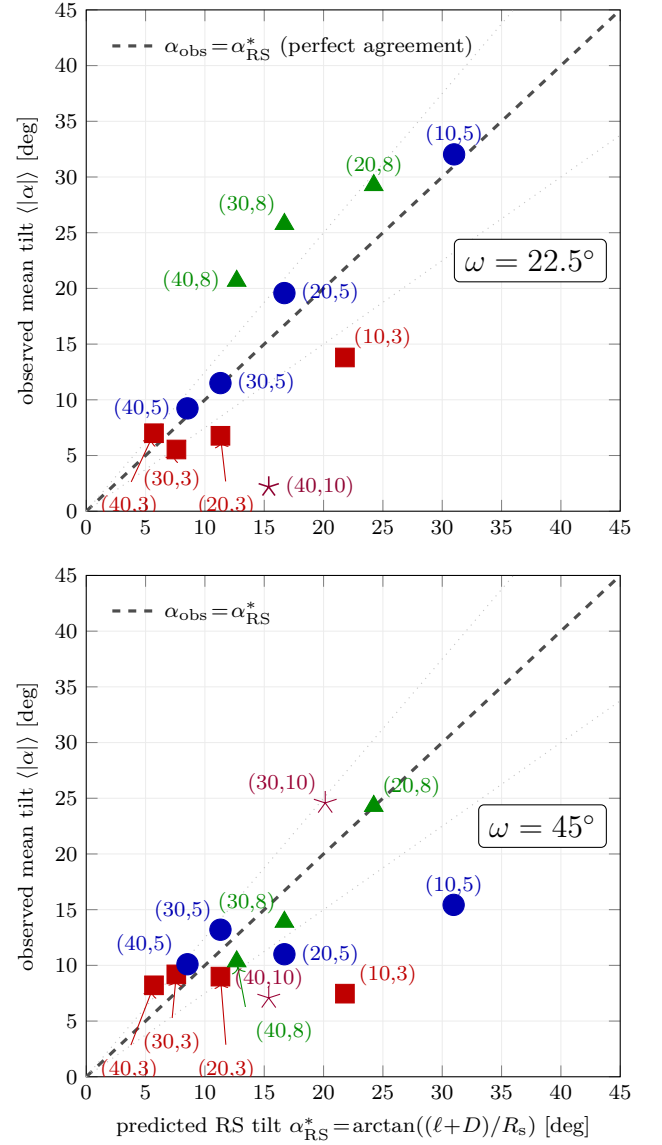
\begin{figure}[H]
\centering
\begin{tikzpicture}
\begin{axis}[
    width=\columnwidth, height=0.95\columnwidth,
    ylabel={observed mean tilt $\langle|\alpha|\rangle$ [deg]},
    xlabel style={font=\footnotesize},
    ylabel style={font=\footnotesize},
    xmin=0, xmax=45,
    ymin=0, ymax=45,
    xtick={0,5,10,15,20,25,30,35,40,45},
    ytick={0,5,10,15,20,25,30,35,40,45},
    xticklabel style={font=\footnotesize},
    yticklabel style={font=\footnotesize},
    grid=both, grid style={gray!15, very thin},
    legend pos=north west,
    legend style={font=\footnotesize, draw=none, fill=none},
  ]
  \addplot[very thick, dashed, gray!60!black, domain=0:45, samples=2]
    {x};
  \addlegendentry{$\alpha_{\rm obs}\!=\!\alpha^{\ast}_{\rm RS}$ (perfect agreement)}
  \addplot[thin, dotted, gray!50, domain=0:45, samples=2] {1.25*x};
  \addplot[thin, dotted, gray!50, domain=0:45, samples=2] {0.75*x};

  \addplot[only marks, mark=square*, mark size=3.5pt, color=red!75!black]
    coordinates {
      (21.80, 13.80)  
      (11.31,  6.77)  
      ( 7.60,  5.54)  
      ( 5.71,  7.00)  
    };

  \addplot[only marks, mark=*, mark size=4pt, color=blue!70!black]
    coordinates {
      (31.00, 32.03)  
      (16.70, 19.58)  
      (11.31, 11.50)  
      ( 8.53,  9.23)  
    };

  \addplot[only marks, mark=triangle*, mark size=4pt, color=green!55!black]
    coordinates {
      (24.23, 29.26)  
      (16.70, 25.77)  
      (12.68, 20.67)  
    };

  \addplot[only marks, mark=star, mark size=4pt, color=purple!75!black,
           line width=0.6pt]
    coordinates {
      (15.38,  2.18)  
    };

  \node[font=\footnotesize, text=blue!70!black, anchor=south]
    at (axis cs:31.00,32.03) {(10,5)};
  \node[font=\footnotesize, text=blue!70!black, anchor=west, xshift=3pt]
    at (axis cs:16.70,19.58) {(20,5)};
  \node[font=\footnotesize, text=blue!70!black, anchor=west, xshift=3pt]
    at (axis cs:11.31,11.50) {(30,5)};
  \node[font=\footnotesize, text=blue!70!black, anchor=east, xshift=-3pt]
    at (axis cs:8.53,9.23) {(40,5)};
  \node[font=\footnotesize, text=green!55!black, anchor=south]
    at (axis cs:24.23,29.26) {(20,8)};
  \node[font=\footnotesize, text=green!55!black, anchor=south east]
    at (axis cs:16.70,25.77) {(30,8)};
  \node[font=\footnotesize, text=green!55!black, anchor=east, xshift=-3pt]
    at (axis cs:12.68,20.67) {(40,8)};
  \node[font=\footnotesize, text=red!75!black, anchor=south west]
    at (axis cs:21.80,13.80) {(10,3)};
  \node[font=\footnotesize, text=red!75!black, anchor=north]
    at (axis cs:11.8,2.2) {(20,3)};
  \draw[red!75!black, thin, ->, shorten <=2pt, shorten >=2pt]
    (axis cs:11.8,2.2) -- (axis cs:11.31,6.77);
  \node[font=\footnotesize, text=red!75!black, anchor=north]
    at (axis cs:7.2,4.6) {(30,3)};
  \draw[red!75!black, thin, ->, shorten <=2pt, shorten >=2pt]
    (axis cs:7.2,4.6) -- (axis cs:7.60,5.54);
  \node[font=\footnotesize, text=red!75!black, anchor=north]
    at (axis cs:3.6,2.2) {(40,3)};
  \draw[red!75!black, thin, ->, shorten <=2pt, shorten >=2pt]
    (axis cs:3.6,2.2) -- (axis cs:5.71,7.00);
  \node[font=\footnotesize, text=purple!75!black, anchor=west, xshift=3pt]
    at (axis cs:15.38,2.18) {(40,10)};
  \node[anchor=east, draw=black, fill=white, rounded corners=2pt,
        inner sep=4pt, font=\large] at (rel axis cs:0.975,0.5)
    {$\tilt=22.5^{\circ}$};
\end{axis}
\end{tikzpicture}
\\[6pt]
\begin{tikzpicture}
\begin{axis}[
    width=\columnwidth, height=0.95\columnwidth,
    xlabel={predicted RS tilt
      $\alpha^{\ast}_{\rm RS}\!=\!\arctan((\Lrod\!+\!\Drod)/\Rsph)$ [deg]},
    ylabel={observed mean tilt $\langle|\alpha|\rangle$ [deg]},
    xlabel style={font=\footnotesize},
    ylabel style={font=\footnotesize},
    xmin=0, xmax=45,
    ymin=0, ymax=45,
    xtick={0,5,10,15,20,25,30,35,40,45},
    ytick={0,5,10,15,20,25,30,35,40,45},
    xticklabel style={font=\footnotesize},
    yticklabel style={font=\footnotesize},
    grid=both, grid style={gray!15, very thin},
    legend pos=north west,
    legend style={font=\footnotesize, draw=none, fill=none},
  ]
  \addplot[very thick, dashed, gray!60!black, domain=0:45, samples=2]
    {x};
  \addlegendentry{$\alpha_{\rm obs}\!=\!\alpha^{\ast}_{\rm RS}$}
  \addplot[thin, dotted, gray!50, domain=0:45, samples=2] {1.25*x};
  \addplot[thin, dotted, gray!50, domain=0:45, samples=2] {0.75*x};

  \addplot[only marks, mark=square*, mark size=3.5pt, color=red!75!black]
    coordinates {
      (21.80, 7.46)   
      (11.31, 8.99)   
      ( 7.59, 9.19)   
      ( 5.71, 8.21)   
    };
  \addplot[only marks, mark=*, mark size=4pt, color=blue!70!black]
    coordinates {
      (30.96, 15.43)  
      (16.70, 11.00)  
      (11.31, 13.20)  
      ( 8.53, 10.10)  
    };
  \addplot[only marks, mark=triangle*, mark size=4pt, color=green!55!black]
    coordinates {
      (24.23, 24.30)  
      (16.70, 13.90)  
      (12.68, 10.35)  
    };
  \addplot[only marks, mark=star, mark size=4.5pt, color=purple!80!black]
    coordinates {
      (20.14, 24.58)  
      (15.38,  7.08)  
    };

  \node[font=\footnotesize, text=blue!70!black, anchor=south]
    at (axis cs:30.96,15.43) {(10,5)};
  \node[font=\footnotesize, text=blue!70!black, anchor=west, xshift=3pt]
    at (axis cs:16.70,11.00) {(20,5)};
  \node[font=\footnotesize, text=blue!70!black, anchor=south east]
    at (axis cs:11.31,13.20) {(30,5)};
  \node[font=\footnotesize, text=blue!70!black, anchor=south east, xshift=-2pt]
    at (axis cs:8.53,10.10) {(40,5)};
  \node[font=\footnotesize, text=green!55!black, anchor=south west]
    at (axis cs:24.23,24.30) {(20,8)};
  \node[font=\footnotesize, text=green!55!black, anchor=south east]
    at (axis cs:16.70,13.90) {(30,8)};
  \node[font=\footnotesize, text=green!55!black, anchor=north west]
    at (axis cs:13.5,5.8) {(40,8)};
  \draw[green!55!black, thin, ->, shorten <=2pt, shorten >=2pt]
    (axis cs:13.5,5.8) -- (axis cs:12.68,10.35);
  \node[font=\footnotesize, text=red!75!black, anchor=south west, xshift=2pt]
    at (axis cs:21.80,7.46) {(10,3)};
  \node[font=\footnotesize, text=red!75!black, anchor=north]
    at (axis cs:11.8,2.4) {(20,3)};
  \draw[red!75!black, thin, ->, shorten <=2pt, shorten >=2pt]
    (axis cs:11.8,2.4) -- (axis cs:11.31,8.99);
  \node[font=\footnotesize, text=red!75!black, anchor=north]
    at (axis cs:7.2,4.8) {(30,3)};
  \draw[red!75!black, thin, ->, shorten <=2pt, shorten >=2pt]
    (axis cs:7.2,4.8) -- (axis cs:7.59,9.19);
  \node[font=\footnotesize, text=red!75!black, anchor=north]
    at (axis cs:3.6,2.4) {(40,3)};
  \draw[red!75!black, thin, ->, shorten <=2pt, shorten >=2pt]
    (axis cs:3.6,2.4) -- (axis cs:5.71,8.21);
  \node[font=\footnotesize, text=purple!80!black, anchor=south east]
    at (axis cs:20.14,24.58) {(30,10)};
  \node[font=\footnotesize, text=purple!80!black, anchor=south, yshift=0pt]
    at (axis cs:15.38,7.08) {(40,10)};
  \node[anchor=east, draw=black, fill=white, rounded corners=2pt,
        inner sep=4pt, font=\large] at (rel axis cs:0.975,0.5)
    {$\tilt=45^{\circ}$};
\end{axis}
\end{tikzpicture}
\caption[Test of the P3 tilt-angle formula]{Master test of the P3 tilt-angle formula
[Eq.\eqref{eq:SP6}]: observed mean layer tilt
$\langle|\alpha|\rangle$ versus the predicted
$\alpha^{\ast}_{\rm RS}=\arctan[(\Lrod+\Drod)/\Rsph]$ for every
$(\Rsph,\Lrod)$ cohort with detectable smectic phase at
$\tilt=22.5^{\circ}$ 
and $\tilt=45^{\circ}$ 
Marker shape and colour encode rod length, 
and  the dashed diagonal is perfect prediction.
The chirality-coherent cohorts ($\chi\ge1.5$, classified in
Fig.\ref{fig:phasediag}), which are the points expected to lie on the
diagonal, are $(10,5)$, $(20,5)$, $(30,5)$, $(20,8)$, $(30,8)$, and
$(40,8)$ at $\tilt=22.5^{\circ}$ (top) and include $(30,10)$ at
 $\tilt=45^{\circ}$ (bottom). The remaining cohorts are Sm-A or
multi-domain and are not expected to track the prediction.
\rNEW{They are retained to show the negative controls rather than as
positive evidence for the tilt scale.}}
\label{fig:heatmaps}
\end{figure}

Figure~\ref{fig:heatmaps} is the central quantitative test of
P3: the measured mean layer tilt against the parameter-free
prediction \eqref{eq:SP6} across the full $(\Rsph,\Lrod)$
landscape, at the two slices  
of $\tilt$.
We make three observations. First, the
chirality-coherent cohorts follow the
diagonal, at $\tilt=22.5^{\circ}$ the six coherent panels
$(10,5)$, $(20,5)$, $(30,5)$, $(20,8)$, $(30,8)$, $(40,8)$ have
measured tilts that scale with $\arctan[(\Lrod+\Drod)/\Rsph]$
with no fitted constant, four of them inside the $\pm 25\%$
band; the two larger-aspect panels $(30,8)$ and $(40,8)$
exceed the prediction by $50$--$65\%$.
Thus, Eq.\eqref{eq:SP6} is a
leading curvature scale accurate at $\Lrod=5\Drod$ but incomplete
for $\Lrod\ge 8\Drod$, where additional aspect-ratio corrections
enlarge the realised tilt.
Second, the noncoherent
cohorts at this $22.5^{\circ}$ slice (all $\Lrod=3\Drod$, plus $(40,5)$, $(40,10)$)
remain near the $\sigma_{0}\sim 8^{\circ}$ noise level.
Third, comparing the two slices $\tilt=22.5^{\circ}$ and $45^{\circ}$ shows the chirality signal
concentrated at low tilt: the signed mean is largest near
$22.5^{\circ}$ and falls toward $45^{\circ}$, so by $45^{\circ}$
every $\Lrod=5\Drod$ cohort has dropped below the coherence
threshold $1.5\,\sigma_{0}$. Only the longest rods stay coherent
there:  $(20,8)$ and $(30,10)$, whose larger tilt magnitude
$|\alpha^{\ast}|$ keeps $|\langle\alpha\rangle|$ above the
threshold.
Read together, the first and third observations describe a
single length--tilt crossover: the rod length that quantitatively
realises the parameter-free tilt angle shifts upward as the imposed director
tilt is raised. At $22.5^{\circ}$ the $\Lrod=5\Drod$ cohorts lie on the
diagonal while the $\Lrod=8\Drod$ cohorts over-tilt, whereas at
$45^{\circ}$ the $\Lrod=8\Drod$ cohort $(20,8)$ relaxes onto the diagonal
(measured $24.3^{\circ}$ versus predicted $24.2^{\circ}$) at the same tilt
where every $\Lrod=5\Drod$ cohort has fallen below the coherence
threshold.

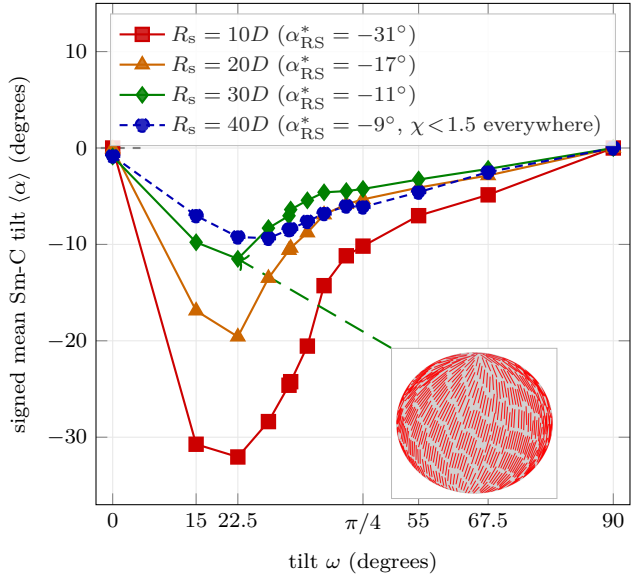
\begin{figure}[H]
\centering
\begin{tikzpicture}
\begin{axis}[
    width=\columnwidth, height=0.95\columnwidth,
    xlabel={tilt $\tilt$ (degrees)},
    ylabel={signed mean Sm-C tilt $\langle\alpha\rangle$ (degrees)},
    xlabel style={font=\footnotesize},
    ylabel style={font=\footnotesize},
    xmin=-3, xmax=93,
    ymin=-37, ymax=15,
    xtick={0,15,22.5,45,55,67.5,90},
    xticklabels={$0$,$15$,$22.5$,$\pi/4$,$55$,$67.5$,$90$},
    xticklabel style={font=\footnotesize},
    yticklabel style={font=\footnotesize},
    grid=both,
    grid style={gray!18},
    legend cell align=left,
    legend style={at={(0.5,0.98)}, anchor=north,
      font=\footnotesize, fill opacity=0.85, draw=gray!50}
]
\addplot[gray, dashed, thick, samples=2, forget plot] {0};

\addplot[mark=square*, mark size=2.5pt, color=red!80!black, thick] coordinates {
  (0.0,     0.0)
  (15.0,  -30.73)
  (22.5,  -32.05)
  (28.0,  -28.38)
  (31.7,  -24.61)
  (32.0,  -24.24)
  (35.0,  -20.55)
  (38.0,  -14.28)
  (42.0,  -11.18)
  (45.0,  -10.18)
  (55.0,   -7.02)
  (67.5,   -4.86)
  (90.0,    0.0)
};
\addlegendentry{$\Rsph=10\Drod$ ($\alpha^{\ast}_{\rm RS}=-31^{\circ}$)}

\addplot[mark=triangle*, mark size=2.8pt, color=orange!80!black, thick] coordinates {
  (0.0,    -0.07)
  (15.0,  -16.88)
  (22.5,  -19.58)
  (28.0,  -13.50)
  (31.7,  -10.60)
  (32.0,  -10.37)
  (35.0,   -8.78)
  (38.0,   -6.91)
  (42.0,   -5.84)
  (45.0,   -5.33)
  (55.0,   -4.11)
  (67.5,   -2.85)
  (90.0,    0.0)
};
\addlegendentry{$\Rsph=20\Drod$ ($\alpha^{\ast}_{\rm RS}=-17^{\circ}$)}

\addplot[mark=diamond*, mark size=2.8pt, color=green!50!black, thick] coordinates {
  (0.0,    -0.88)
  (15.0,   -9.78)
  (22.5,  -11.50)
  (28.0,   -8.31)
  (31.7,   -7.03)
  (32.0,   -6.39)
  (35.0,   -5.43)
  (38.0,   -4.60)
  (42.0,   -4.47)
  (45.0,   -4.25)
  (55.0,   -3.27)
  (67.5,   -2.17)
  (90.0,    0.0)
};
\addlegendentry{$\Rsph=30\Drod$ ($\alpha^{\ast}_{\rm RS}=-11^{\circ}$)}

\addplot[mark=*, mark size=2.5pt, color=blue!70!black, thick,
         dash pattern=on 3pt off 2pt] coordinates {
  (0.0,    -0.83)
  (15.0,   -7.02)
  (22.5,   -9.23)
  (28.0,   -9.37)
  (31.7,   -8.47)
  (32.0,   -8.40)
  (35.0,   -7.62)
  (38.0,   -6.80)
  (42.0,   -6.02)
  (45.0,   -6.12)
  (55.0,   -4.58)
  (67.5,   -2.50)
  (90.0,    0.0)
};
\addlegendentry{$\Rsph=40\Drod$ ($\alpha^{\ast}_{\rm RS}=-9^{\circ}$, $\chi\!<\!1.5$ everywhere)}

\node[fill=white, draw=gray!55, line width=0.3pt, inner sep=1pt,
      anchor=south] (smcsnap) at (axis cs:65,-36.4)
  {\includegraphics[width=0.245\columnwidth]{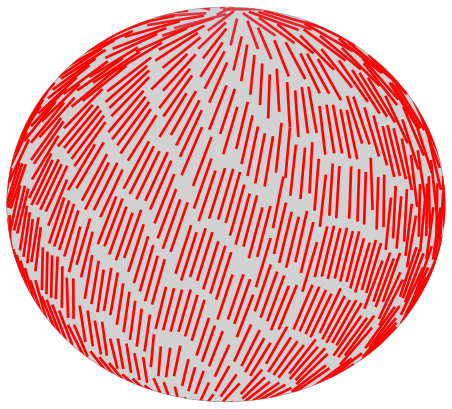}};
\draw[->, thick, green!50!black, dash pattern=on 8pt off 4pt]
  (smcsnap.north west) -- (axis cs:22.5,-11.5);

\end{axis}
\end{tikzpicture}
\caption{Sm-C signed-mean tilt versus $\tilt$ across host radii
($\Lrod=5\Drod$). Signed mean Sm-C tilt
$\langle\alpha\rangle(\tilt)$ for the four host radii; the legend
lists the P3 prediction $-|\alpha^{\ast}|$ of
Eq.\eqref{eq:SP6} for each radius. The dashed horizontal line is
the Sm-A reference $\langle\alpha\rangle=0$; the $\Rsph=40\Drod$
trace is shown dashed. The magnitude divided by the per-host
width $\sigma_{\alpha}$ gives the chirality index of
Fig.\ref{fig:chirality}. Inset (lower left): a representative
coherent Sm-C texture; the long-dashed arrow marks the
$\tilt=22.5^{\circ}$ minimum of the $\Rsph=30\Drod$ trace.}
\label{fig:phasediagL5}
\end{figure}

\begin{figure}[H]
\centering
\begin{tikzpicture}
\begin{axis}[
    width=\columnwidth, height=0.78\columnwidth,
    xlabel={tilt $\tilt$ (degrees)},
    ylabel={chirality index $|\langle\alpha\rangle|/\sigma_{\alpha}$},
    xlabel style={font=\footnotesize},
    ylabel style={font=\footnotesize},
    xmin=-3, xmax=93,
    ymin=0, ymax=4.0,
    xtick={0,15,22.5,45,55,67.5,90},
    xticklabels={$0$,$15$,$22.5$,$\pi/4$,$55$,$67.5$,$90$},
    xticklabel style={font=\footnotesize},
    yticklabel style={font=\footnotesize},
    grid=both,
    grid style={gray!18},
    legend pos=north east,
    legend cell align=left,
    legend style={font=\footnotesize,fill opacity=0.85,draw=gray!50}
]
\addplot[gray, dashed, thick, samples=2, forget plot] {1.5};
\addplot[gray, dotted, thick, samples=2, forget plot] {1.0};

\addplot[mark=square*, mark size=2.5pt, color=red!80!black, thick] coordinates {
  (0.0,    0.0)
  (15.0,   3.5)
  (22.5,   3.65)
  (28.0,   3.2318339)
  (31.7,   2.8027682)
  (32.0,   2.7612457)
  (35.0,   2.34)
  (38.0,   1.6262976)
  (42.0,   1.2733564)
  (45.0,   1.16)
  (55.0,   0.8)
  (67.5,   0.55363322)
  (90.0,   0.0)
};
\addlegendentry{$\Rsph=10\Drod$}

\addplot[mark=triangle*, mark size=2.8pt, color=orange!80!black, thick] coordinates {
  (0.0,    0.01)
  (15.0,   2.5)
  (22.5,   2.9)
  (28.0,   2.0)
  (31.7,   1.57)
  (32.0,   1.5363322)
  (35.0,   1.3010381)
  (38.0,   1.0242215)
  (42.0,   0.8650519)
  (45.0,   0.78892734)
  (55.0,   0.60899654)
  (67.5,   0.42214533)
  (90.0,   0.0)
};
\addlegendentry{$\Rsph=20\Drod$}

\addplot[mark=diamond*, mark size=2.8pt, color=green!50!black, thick] coordinates {
  (0.0,    0.1384083)
  (15.0,   1.53)
  (22.5,   1.80)
  (28.0,   1.30)
  (31.7,   1.10)
  (32.0,   1.00)
  (35.0,   0.85)
  (38.0,   0.72)
  (42.0,   0.69896194)
  (45.0,   0.66435986)
  (55.0,   0.51211073)
  (67.5,   0.34)
  (90.0,   0.0)
};
\addlegendentry{$\Rsph=30\Drod$}

\addplot[mark=*, mark size=2.5pt, color=blue!70!black, thick] coordinates {
  (0.0,    0.083044983)
  (15.0,   0.7)
  (22.5,   0.92)
  (28.0,   0.93425606)
  (31.7,   0.84429066)
  (32.0,   0.83737024)
  (35.0,   0.76)
  (38.0,   0.67820069)
  (42.0,   0.60)
  (45.0,   0.61)
  (55.0,   0.4567474)
  (67.5,   0.24913495)
  (90.0,   0.0)
};
\addlegendentry{$\Rsph=40\Drod$}
\end{axis}
\end{tikzpicture}
\caption{Sm-C global chirality versus tilt $\tilt$ across host
radii ($\Lrod=5\Drod$): the chirality index
$\chi=|\langle\alpha\rangle|/\sigma_{\alpha}$ for the same
panels as Fig.\ref{fig:phasediagL5}. The horizontal dashed line
marks the coherent-Sm-C threshold $\chi=1.5$: a host carries one
coherent Sm-C handedness at tilts where its curve lies above the
line, and is Sm-A or multi-domain where it lies below.
\rNEW{The dotted line marks $\chi=1$, the signed-tilt/noise scale used as
a visual reference.}}
\label{fig:chirality}
\end{figure}
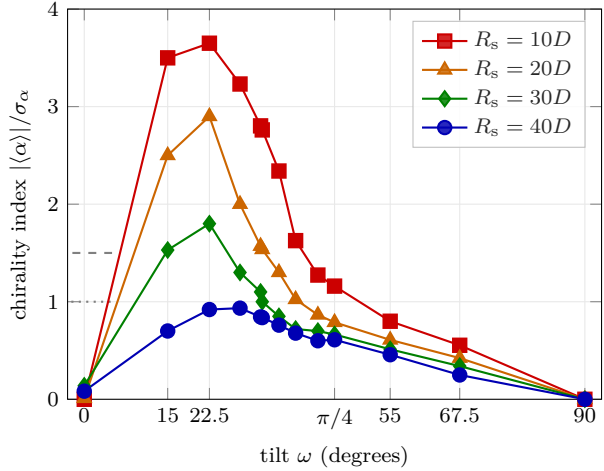

Figures~\ref{fig:phasediagL5} and \ref{fig:chirality} resolve the
$\Lrod=5\Drod$ series in tilt. The signed mean tilt
(Fig.\ref{fig:phasediagL5}) is negative---one global
handedness---over almost the entire interior tilt range,
with peak depth ordered by host radius exactly as
Eq.\eqref{eq:SP6} orders $|\alpha^{\ast}|$: the peak depth is
$-32^{\circ}$ measured versus the predicted tilt $|\alpha^{\ast}|=-31^{\circ}$ at
$\Rsph=10\Drod$, $-20^{\circ}$ versus
$-17^{\circ}$ at $20\Drod$, and progressively shallower traces
at $30$ and $40\Drod$. The peak sits at $\tilt\approx22.5^{\circ}$
(at $28^{\circ}$ on the shallow $40\Drod$ trace), below
the symmetric point $\pi/4$, with the envelope concentrated on the
low-tilt side, matching the low-tilt peak
$\tilt_{\rm pk}\approx23.6^{\circ}$ of the calibrated
($\phigold$-consistent) envelope [Eqs.~\eqref{eq:shift},~\eqref{eq:peak}].
The
chirality index (Fig.\ref{fig:chirality}) measures the
\emph{signed} mean tilt relative to its scatter, so it is
nonzero only when the whole sphere selects one handedness: a
local layer tilt is necessary but not sufficient, and the
chirality index carries the phase assignment. In
Fig.\ref{fig:chirality} the two smallest hosts $\Rsph=10$ and
$20\Drod$ cross the $\chi\ge 1.5$ threshold over finite tilt
windows, peaking at $\chi\approx 3.65$ and $\chi\approx 2.9$ near
$\tilt=22.5^{\circ}$.  $\Rsph=30\Drod$ crosses over a narrow
low-tilt window ($\chi\approx 1.53$ at $15^{\circ}$ and
$\chi\approx 1.8$ at $22.5^{\circ}$), and $\Rsph=40\Drod$ never
does ($\chi\le 1.0$). The observed coherence edge therefore lies
between $30$ and $40\Drod$, matching the parameter-free geometric
boundary $R^{\ast}_{\rm SmC}(22.5^{\circ})\approx5(\Lrod+\Drod)=30\Drod$
[Eq.\eqref{eq:SP5}] for $\Lrod=5\Drod$, with no $\sigma_{0}$
calibration.

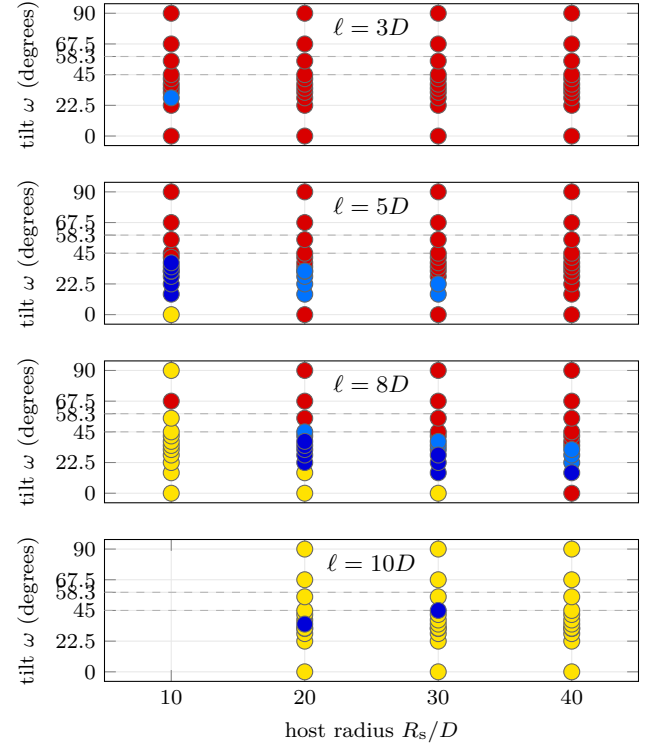
\begin{figure}[H]
\centering
\begin{tikzpicture}
\begin{axis}[
    width=\columnwidth, height=0.40\columnwidth,
    ylabel={tilt $\tilt$ (degrees)},
    ylabel style={font=\footnotesize},
    xmin=5, xmax=45,
    ymin=-7, ymax=97,
    xtick={10,20,30,40},
    xticklabels={},
    ytick={0,22.5,45,58.3,67.5,90},
    yticklabel style={font=\footnotesize},
    grid=both, grid style={gray!18},
]
\addplot[only marks, mark=*, mark size=3pt, color=red!85!black, draw=black!60, fill=red!85!black] coordinates {
  (10, 0.0) (10, 22.5) (10, 32.0) (10, 35.0) (10, 38.0) (10, 42.0)
  (10, 45.0) (10, 55.0) (10, 67.5) (10, 90.0)
  (20, 0.0) (20, 22.5) (20, 28.0) (20, 32.0) (20, 35.0) (20, 38.0)
  (20, 42.0) (20, 45.0) (20, 55.0) (20, 67.5) (20, 90.0)
  (30, 0.0) (30, 22.5) (30, 28.0) (30, 32.0) (30, 35.0) (30, 38.0)
  (30, 42.0) (30, 45.0) (30, 55.0) (30, 67.5) (30, 90.0)
  (40, 0.0) (40, 22.5) (40, 28.0) (40, 32.0) (40, 35.0) (40, 38.0)
  (40, 42.0) (40, 45.0) (40, 55.0) (40, 67.5) (40, 90.0)
};
\addplot[only marks, mark=*, mark size=3pt, color=blue!55!cyan, draw=black!60, fill=blue!55!cyan] coordinates {
  (10, 28.0)
};
\draw[dashed, gray!60, line width=0.4pt]
    ({rel axis cs:0,0}|-{axis cs:5,45}) -- ({rel axis cs:1,0}|-{axis cs:5,45});
\draw[dashed, gray!60, line width=0.4pt]
    ({rel axis cs:0,0}|-{axis cs:5,58.3}) -- ({rel axis cs:1,0}|-{axis cs:5,58.3});
\node[font=\small] at (axis cs:25, 80) {$\Lrod=3\Drod$};
\end{axis}
\end{tikzpicture}
\\[2pt]
\begin{tikzpicture}
\begin{axis}[
    width=\columnwidth, height=0.40\columnwidth,
    ylabel={tilt $\tilt$ (degrees)},
    ylabel style={font=\footnotesize},
    xmin=5, xmax=45,
    ymin=-7, ymax=97,
    xtick={10,20,30,40},
    xticklabels={},
    ytick={0,22.5,45,58.3,67.5,90},
    yticklabel style={font=\footnotesize},
    grid=both, grid style={gray!18},
]
\addplot[only marks, mark=*, mark size=3pt, color=yellow!85!orange, draw=black!60, fill=yellow!85!orange] coordinates {
  (10, 0.0)
};
\addplot[only marks, mark=*, mark size=3pt, color=red!85!black, draw=black!60, fill=red!85!black] coordinates {
  (40, 15.0) (10, 45.0) (10, 42.0) (10, 55.0) (10, 67.5) (10, 90.0)
  (20, 0.0) (20, 35.0) (20, 38.0) (20, 42.0) (20, 45.0) (20, 55.0)
  (20, 67.5) (20, 90.0)
  (30, 0.0) (30, 28.0) (30, 32.0) (30, 35.0) (30, 38.0) (30, 42.0)
  (30, 45.0) (30, 55.0) (30, 67.5) (30, 90.0)
  (40, 0.0) (40, 22.5) (40, 28.0) (40, 32.0) (40, 35.0) (40, 38.0)
  (40, 42.0) (40, 45.0) (40, 55.0) (40, 67.5) (40, 90.0)
};
\addplot[only marks, mark=*, mark size=3pt, color=blue!55!cyan, draw=black!60, fill=blue!55!cyan] coordinates {
  (20, 15.0) (20, 22.5) (20, 28.0) (20, 31.7) (20, 32.0)
  (30, 15.0) (30, 22.5)
};
\addplot[only marks, mark=*, mark size=3pt, color=blue!85!black, draw=black!60, fill=blue!85!black] coordinates {
  (10, 15.0) (10, 22.5) (10, 28.0) (10, 31.7) (10, 32.0) (10, 35.0) (10, 38.0)
};
\draw[dashed, gray!60, line width=0.4pt]
    ({rel axis cs:0,0}|-{axis cs:5,45}) -- ({rel axis cs:1,0}|-{axis cs:5,45});
\draw[dashed, gray!60, line width=0.4pt]
    ({rel axis cs:0,0}|-{axis cs:5,58.3}) -- ({rel axis cs:1,0}|-{axis cs:5,58.3});
\node[font=\small] at (axis cs:25, 80) {$\Lrod=5\Drod$};
\end{axis}
\end{tikzpicture}
\\[2pt]
\begin{tikzpicture}
\begin{axis}[
    width=\columnwidth, height=0.40\columnwidth,
    ylabel={tilt $\tilt$ (degrees)},
    ylabel style={font=\footnotesize},
    xmin=5, xmax=45,
    ymin=-7, ymax=97,
    xtick={10,20,30,40},
    xticklabels={},
    ytick={0,22.5,45,58.3,67.5,90},
    yticklabel style={font=\footnotesize},
    grid=both, grid style={gray!18},
]
\addplot[only marks, mark=*, mark size=3pt, color=yellow!85!orange, draw=black!60, fill=yellow!85!orange] coordinates {
  (10, 0.0) (10, 15.0) (20, 15.0) (10, 22.5) (10, 28.0) (10, 32.0)
  (10, 35.0) (10, 38.0) (10, 42.0) (10, 45.0) (10, 55.0) (10, 90.0)
  (20, 0.0) (30, 0.0)
};
\addplot[only marks, mark=*, mark size=3pt, color=red!85!black, draw=black!60, fill=red!85!black] coordinates {
  (10, 67.5) (20, 55.0) (20, 67.5) (20, 90.0)
  (30, 42.0) (30, 45.0) (30, 55.0) (30, 67.5) (30, 90.0)
  (40, 0.0) (40, 35.0) (40, 38.0) (40, 42.0) (40, 45.0) (40, 55.0)
  (40, 67.5) (40, 90.0)
};
\addplot[only marks, mark=*, mark size=3pt, color=blue!55!cyan, draw=black!60, fill=blue!55!cyan] coordinates {
  (20, 42.0) (20, 45.0) (30, 32.0) (30, 35.0) (30, 38.0)
  (40, 22.5) (40, 28.0) (40, 32.0)
};
\addplot[only marks, mark=*, mark size=3pt, color=blue!85!black, draw=black!60, fill=blue!85!black] coordinates {
  (30, 15.0) (40, 15.0) (20, 22.5) (20, 28.0) (20, 32.0) (20, 35.0)
  (20, 38.0) (30, 22.5) (30, 28.0)
};
\draw[dashed, gray!60, line width=0.4pt]
    ({rel axis cs:0,0}|-{axis cs:5,45}) -- ({rel axis cs:1,0}|-{axis cs:5,45});
\draw[dashed, gray!60, line width=0.4pt]
    ({rel axis cs:0,0}|-{axis cs:5,58.3}) -- ({rel axis cs:1,0}|-{axis cs:5,58.3});
\node[font=\small] at (axis cs:25, 80) {$\Lrod=8\Drod$};
\end{axis}
\end{tikzpicture}
\\[2pt]
\begin{tikzpicture}
\begin{axis}[
    width=\columnwidth, height=0.40\columnwidth,
    xlabel={host radius $\Rsph/\Drod$},
    ylabel={tilt $\tilt$ (degrees)},
    xlabel style={font=\footnotesize},
    ylabel style={font=\footnotesize},
    xmin=5, xmax=45,
    ymin=-7, ymax=97,
    xtick={10,20,30,40},
    ytick={0,22.5,45,58.3,67.5,90},
    xticklabel style={font=\footnotesize},
    yticklabel style={font=\footnotesize},
    grid=both, grid style={gray!18},
]
\addplot[only marks, mark=*, mark size=3pt, color=yellow!85!orange, draw=black!60, fill=yellow!85!orange] coordinates {
  (20, 0.0) (20, 22.5) (20, 28.0) (20, 31.7) (20, 32.0) (20, 38.0)
  (20, 42.0) (20, 45.0) (20, 55.0) (20, 67.5) (20, 90.0)
  (30, 0.0) (30, 22.5) (30, 28.0) (30, 31.7) (30, 32.0) (30, 35.0)
  (30, 38.0) (30, 42.0) (30, 55.0) (30, 67.5) (30, 90.0)
  (40, 0.0) (40, 22.5) (40, 28.0) (40, 31.7) (40, 32.0) (40, 35.0)
  (40, 38.0) (40, 42.0) (40, 45.0) (40, 55.0) (40, 67.5) (40, 90.0)
};
\addplot[only marks, mark=*, mark size=3pt, color=blue!85!black, draw=black!60, fill=blue!85!black] coordinates {
  (20, 35.0) (30, 45.0)
};
\draw[dashed, gray!60, line width=0.4pt]
    ({rel axis cs:0,0}|-{axis cs:5,45}) -- ({rel axis cs:1,0}|-{axis cs:5,45});
\draw[dashed, gray!60, line width=0.4pt]
    ({rel axis cs:0,0}|-{axis cs:5,58.3}) -- ({rel axis cs:1,0}|-{axis cs:5,58.3});
\node[font=\small] at (axis cs:25, 80) {$\Lrod=10\Drod$};
\end{axis}
\end{tikzpicture}
\caption{Two-dimensional Sm-A/Sm-C phase diagrams on the
$(\Rsph,\tilt)$ plane for $\Lrod=3\Drod$ (top),
$5\Drod$, $8\Drod$, and $10\Drod$ (bottom). Each marker classifies
one strict-MC panel by the chirality index:
\legball{yellow!85!orange}~nematic ($\chi_{\rm sm}<0.10$);
\legball{red!85!black}~Sm-A / noncoherent ($\chi<1.5$);
\legball{blue!55!cyan}~coherent Sm-C with
$|\langle\alpha\rangle|\in(12^{\circ},22^{\circ}]$;
\legball{blue!85!black}~coherent Sm-C with
$|\langle\alpha\rangle|>22^{\circ}$.
The $\Lrod=10\Drod$ panel has no
$\Rsph=10\Drod$ column which is geometrically
frustrated ($\Lrod/\Rsph=1$. It is above the smectic-formation bound
$\Lrod/\Rsph\sim 1/3$) and thus was not simulated.
}
\label{fig:phasediag}
\end{figure}

The two-dimensional phase diagrams
(Fig.\ref{fig:phasediag}) assemble the
panel-by-panel classification into the $(\Rsph,\tilt)$ plane and
resolve three length-dependent regimes. For the shortest rods
($\Lrod=3\Drod$) the cohorts barely sense the host curvature and
behave like rods on a nearly flat substrate: the layers remain
Sm-A, and a single coherent Sm-C panel survives only on the
smallest host ($\Rsph=10\Drod$) at low tilt, where curvature is
strongest. At the opposite extreme ($\Lrod=10\Drod$) the aspect
ratio exceeds the critical value for smectic order, so most panels
register neither coherent Sm-A nor Sm-C and only two isolated Sm-C
points remain. Between these limits ($\Lrod=5$--$8\Drod$) the
simulations detect a robust coherent Sm-C window. In every diagram
this window sits at \emph{low} $\tilt$, consistent with the
monotonic decrease of $R^{\ast}_{\rm SmC}(\tilt)$
[Eq.\eqref{eq:SP5}] and with the chirality envelope
$\sin(2\tilt)\cos^{\phigold^{3}}\!\tilt$ [Eq.\eqref{eq:shift}],
which concentrates the signed mean on the same low-tilt side.

\section{Discussion}\vspace{-0.4cm}

In summary, we have shown that spherical confinement induces a
Sm-C phase in a hard-rod fluid whose bulk phase diagram contains
none, in the strict locked-director limit, and that a single
ratio-symmetric cost functional [Eq.\eqref{eq:Jdef}] organises the
curvature-induced behaviour: the layer spacing $d^{\ast}$, the
smectic-area window [Eq.\eqref{eq:SP4}], the Sm-A\,$\to$\,Sm-C
boundary [Eq.\eqref{eq:SP5}], and the tilt-angle magnitude, low-tilt
sign envelope, and coherence window
[Eqs.~\eqref{eq:SP6}, \eqref{eq:shift}, and~\eqref{eq:Rcoh}], where
the envelope exponent is calibrated to the data and is therefore
$\phigold$-consistent rather than parameter-free.
The load-bearing result is the closed-form boundary
$R^{\ast}_{\rm SmC}(\tilt)$: the simulated host radii lie on both
sides of it, so the data test its location in $\Rsph$, not merely its
proportionality to $\Lrod+\Drod$. Decreasing monotonically with tilt
and equal to $\approx5(\Lrod+\Drod)$ near the peak tilt, this
boundary fixes both the low-tilt location and the radial scale
($\approx30\Drod$ for $\Lrod=5\Drod$) of the coherent block, while
the measured noise floor $\sigma_{0}\approx8^{\circ}$ and the
$\chi\!\ge\!1.5$ detection threshold enter only the coherence test,
not the radius. Its lower edge follows from the reciprocal symmetry of
$\Jcost$ alone, whereas the upper edge $\arctan\phigold=58.3^{\circ}$
remains a channel-saturation hypothesis, and the tilt magnitude
$\arctan[(\Lrod+\Drod)/\Rsph]$ is a leading scale established here
only for the shorter rods.

The natural next step is to relax the locked-orientation constraint
$K_{3}/K_{1}\to\infty$, in which every rod axis is pinned to the
imposed director. Relaxation restores the orientational entropy the
lock suppresses, and the single-handed chirality is the observable
most exposed to it; how the ratio $K_{3}/K_{1}$ regulates that
handedness is the central question it opens. We expect the window
edges, and especially $R^{\ast}_{\rm SmC}(\tilt)$, to survive partial
relaxation, since that radius follows from the geometric cost balance
and curvature rather than from the suppressed entropy. The present
NVT data cannot by themselves decide whether the curvature-induced
Sm-C is thermodynamically stable or a long-lived metastable state, so
whether the coherent handedness persists at finite $K_{3}/K_{1}$
is the decisive open test.

\section*{SUPPLEMENTARY MATERIAL} \vspace{-0.4cm}
See the SM for the recognition-cost derivations,
simulation metadata, and analysis conventions 
supporting the Communication.


\section*{AUTHOR DECLARATIONS} \vspace{-0.4cm}
\subsection*{Conflict of Interest} \vspace{-0.4cm}
The authors have no conflicts to disclose.
\subsection*{Author Contributions} \vspace{-0.4cm}
\textbf{Jonathan Washburn}: Conceptualization (equal);
Methodology (equal); Formal analysis (supporting); Writing --
review \& editing (equal).
\textbf{Hartmut L\"owen}: Conceptualization (equal); Formal
analysis (supporting); Writing -- review \& editing (equal).
\textbf{Elshad Allahyarov}: Conceptualization (equal);
Investigation (lead); Methodology (equal); Software (lead);
Formal analysis (lead); Visualization (lead); Writing -- original
draft (lead); Writing -- review \& editing (equal).

\section*{DATA AVAILABILITY} \vspace{-0.4cm}
The full configuration ensemble, the analysis pipeline, and the
data that support the findings of this study are available from
the corresponding author upon request.

\FloatBarrier

\makeatletter
\immediate\write\@auxout{\string\citation{aip41Control}}
\makeatother
\makeatletter
\let\RSorig@href\href
\def\RS@lastdoi{}
\renewcommand{\href}[2]{\def\RS@lastdoi{#1}\RSorig@href{#1}{#2}}
\providecommand\BibitemShut[1]{}
\def\BibitemShut#1{%
  \csname bibitem#1\endcsname
  \ifx\RS@lastdoi\@empty\else
    \nobreakspace DOI:\nobreakspace\RSorig@href{\RS@lastdoi}{\nolinkurl{\RS@lastdoi}}%
  \fi
  \let\RS@lastdoi\@empty}
\makeatother
%

\clearpage
\setcounter{section}{0}
\setcounter{equation}{0}
\setcounter{figure}{0}
\setcounter{table}{0}
\renewcommand{\thesection}{S\arabic{section}}
\renewcommand{\theequation}{S\arabic{equation}}
\renewcommand{\thefigure}{S\arabic{figure}}
\renewcommand{\thetable}{S\Roman{table}}

\onecolumngrid
\begin{center}
{\large\bfseries Supplementary Material}\\[4pt]
{\large Curvature-induced smectic-C order of tangentially anchored hard
spherocylinders on a sphere with a rigidly locked director field}\\[6pt]
Jonathan Washburn, Hartmut L\"owen, and Elshad Allahyarov
\end{center}
\vspace{4pt}
\twocolumngrid

\noindent
This Supplementary Material gives the algebraic and numerical details that
support the Communication. Section~\ref{sec:si_theory} records the
recognition-cost definitions and the derivations of the geometric factors
used in the main text. Section~\ref{sec:si_methods} records the simulation
metadata and analysis conventions used for the locked-orientation Monte
Carlo data set.

\section{Geometric and recognition-cost derivations}
\label{sec:si_theory}

The locked director field used throughout the Communication is
\begin{equation}
 \nhat_{\tilt}(\theta,\phi)
 =\cos\tilt\,\eyhat+\sin\tilt\,\ephihat ,
 \label{eq:si_director}
\end{equation}
where $\tilt$ is the prescribed tilt from the meridian direction.  The
recognition cost assigned to a scale ratio $x>0$ is
\begin{equation}
 \Jcost(x)=\frac{1}{2}\left(x+\frac{1}{x}\right)-1 .
 \label{eq:si_J}
\end{equation}
The calculations below use only the matched-scale condition $\Jcost(1)=0$,
the near-match quadratic behaviour, and the reciprocal symmetry
$\Jcost(x)=\Jcost(1/x)$, except where the channel-saturation hypothesis is
explicitly invoked.
More explicitly, the working assumptions are: (J1) matched scales carry no
cost; (J2) reciprocal mismatch ratios are equivalent; (J3) the near-match
cost is quadratic; and (J4) saturated recognition channels select ratios
associated with the golden-ratio fixed point. Conditions (J1)--(J3) set the
ratio-symmetric cost balance. Condition (J4) is an additional modelling
hypothesis and is used only for the upper edge of the smectic-area window.

\subsection{Squared channel weights for the smectic-area window}

Resolving Eq.~\eqref{eq:si_director} onto the latitude and meridian
directions gives projection amplitudes $\sin\tilt$ and $\cos\tilt$.
Because the near-match contribution is quadratic, the two
layer-breaking channel weights used in the smectic-area estimate are the
squared projections,
\begin{equation}
 w_{\mathrm A}=\sin^2\tilt,\qquad
 w_{\mathrm B}=\cos^2\tilt,\qquad
 w_{\mathrm A}+w_{\mathrm B}=1 .
 \label{eq:si_weights}
\end{equation}
The two-channel coherence cost used in the Communication is therefore
\begin{equation}
 \mathcal{C}_{\rm sm}(\tilt)
 =w_{\mathrm A}\,\Jcost\!\left(\phigold\,y_{\mathrm A}\right)
 +w_{\mathrm B}\,\Jcost\!\left(\phigold\,y_{\mathrm B}\right),
 \label{eq:si_Csm}
\end{equation}
with mismatch ratios $y_{\mathrm A}\propto 1/\sin\tilt$ and
$y_{\mathrm B}\propto 1/\cos\tilt$. Equation~\eqref{eq:si_Csm} identifies
the competing channels; it is not by itself a closed minimisation
problem because the proportionality constant in the mismatch ratios is
not fixed independently.

The lower edge of the window is the symmetric point
$w_{\mathrm A}=w_{\mathrm B}$, hence $\tilt=\pi/4$. The upper edge uses the
additional channel-saturation hypothesis
\begin{equation}
 \frac{w_{\mathrm B}}{w_{\mathrm A}}
 =\cot^2\tilt^\ast_{\rm sm}
 =\frac{1}{\phigold^2},
 \qquad
 \tan\tilt^\ast_{\rm sm}=\phigold ,
 \label{eq:si_saturation}
\end{equation}
giving $\tilt^\ast_{\rm sm}=\arctan\phigold=58.3^\circ$.

\subsection{Sm-A to Sm-C boundary}

For a tilted Sm-A layer normal that spirals around a latitude circle, the
number of natural layers around the loop is
\begin{equation}
 N(\tilt)=\frac{2\pi\Rsph\sin\tilt}{\Lrod+\Drod}.
 \label{eq:si_loop_count}
\end{equation}
Sharing one residual uncommensurate layer over those $N$ layers gives
the fractional mismatch
\begin{equation}
 \epsilon_{\rm disl}
 =\frac{1}{N(\tilt)}
 =\frac{\Lrod+\Drod}{2\pi\Rsph\sin\tilt},
 \label{eq:si_epsilon}
\end{equation}
and hence the Sm-A spiral-dislocation cost
\begin{equation}
 \Jcost_{\rm disl}(\tilt)
 =\Jcost\!\left(1+\frac{\Lrod+\Drod}
 {2\pi\Rsph\sin\tilt}\right).
 \label{eq:si_Jdisl}
\end{equation}
For the closed-latitude Sm-C geometry, a rod tilted by $\tilt$ projects a
fraction $\cos\tilt$ of its end-to-end length onto the layer normal, so
the uniform tilt-projection cost is
\begin{equation}
 \Jcost_{\rm tilt}(\tilt)=\Jcost(\cos\tilt).
 \label{eq:si_Jtilt}
\end{equation}
The boundary is the equality of these two per-rod costs. Using the
reciprocal symmetry of $\Jcost$,
\begin{equation}
\begin{aligned}
\Jcost_{\rm disl}=\Jcost_{\rm tilt}
&\Longleftrightarrow
\Jcost\!\left(1+\frac{\Lrod+\Drod}{2\pi\Rsph\sin\tilt}\right)
=\Jcost(\cos\tilt) \\
&\Longleftrightarrow
1+\frac{\Lrod+\Drod}{2\pi\Rsph\sin\tilt}
=\frac{1}{\cos\tilt}
=\sec\tilt \\
&\Longleftrightarrow
\Rsph=\frac{\Lrod+\Drod}{2\pi\sin\tilt\,(\sec\tilt-1)} .
\end{aligned}
\label{eq:si_SP5_balance}
\end{equation}
Thus
\begin{equation}
 R^\ast_{\rm SmC}(\tilt)
 =\frac{\Lrod+\Drod}{2\pi\sin\tilt\,(\sec\tilt-1)} .
 \label{eq:si_SP5}
\end{equation}
The factor $\sin\tilt(\sec\tilt-1)$ is therefore derived here from the
geometric/RS cost balance. It is not a published Frank--de~Gennes result.
Standard smectic elasticity motivates a curvature-induced crossover, but
Eq.~\eqref{eq:si_SP5} is the closed-form boundary used in the
Communication.
In a conventional Landau--de~Gennes or smectic-elastic
description~\cite{deGennesProst}, layers
on a sphere carry topological and bending frustration, while rod tilt
relieves part of that frustration at the cost of a tilt penalty. Such a
balance can motivate a crossover radius proportional to
$(\Lrod+\Drod)$, but its absolute location would depend on elastic
coefficients such as splay, bend, compression, and tilt susceptibilities.
Those coefficients are not introduced in the locked-orientation RS model.
Equation~\eqref{eq:si_SP5} should therefore be read as a specific
geometric cost-balance prediction for the frozen-director simulations, not
as a replacement derivation of the Frank free energy.

\subsection{Peak of the calibrated chirality envelope}

The signed chirality envelope in the Communication is
\begin{equation}
 \langle\alpha\rangle(\tilt)\propto
 \pm|\alpha^\ast|\sin(2\tilt)\cos^\nu\tilt .
 \label{eq:si_envelope}
\end{equation}
For arbitrary positive exponent $\nu$,
\begin{equation}
\begin{aligned}
\sin(2\tilt)\cos^\nu\tilt
&=2\sin\tilt\,\cos^{1+\nu}\tilt,\\
\frac{d}{d\tilt}
\left[\sin\tilt\,\cos^{1+\nu}\tilt\right]=0
&\Longrightarrow
\tan^2\tilt_{\rm pk}=\frac{1}{1+\nu}.
\end{aligned}
\label{eq:si_peak_condition}
\end{equation}
For $\nu=\phigold^3$, $1+\phigold^3=2\phigold^2$, and therefore
\begin{equation}
 \tilt_{\rm pk}
 =\arctan\frac{1}{\phigold\sqrt{2}}
 \approx 23.6^\circ .
 \label{eq:si_peak_phi3}
\end{equation}
This exponent is calibrated to place the envelope maximum near the
observed low-tilt maximum; it is $\phigold$-consistent, not
$\phigold$-essential.
Evaluating Eq.~\eqref{eq:si_SP5} near this peak gives the
coherent-block scale
\begin{equation}
 R^\ast_{\rm coh}\equiv R^\ast_{\rm SmC}(\tilt_{\rm pk})
 \approx 5(\Lrod+\Drod),
 \label{eq:si_Rcoh}
\end{equation}
which is a geometric scale for the radial edge, not the operational
definition of the phase. The classification itself is made from the
measured chirality index.

\section{Simulation and analysis details}
\label{sec:si_methods}

\subsection{Locked-orientation Monte Carlo ensemble}

All panels are strict locked-orientation NVT Monte Carlo simulations at
packing fraction $\eta\simeq0.75$, with overlap rejection checked by the
Vega--Lago shortest-distance algorithm.\cite{VegaLago1994} The
rod-length-growth (L-grow) preparation for jamming-limited dense panels
follows the same practical idea as the length-growth procedures used in
earlier spherical-rod simulations.\cite{Allahyarov2017,Allahyarov2018}

\begin{table}[h]
\caption{Simulation sizes and production metadata for the
$\eta\simeq0.75$ locked-orientation data set. The frustrated
$(\Rsph,\Lrod)=(10,10)D$ corner was not simulated.}
\label{tab:si_counts}
\begin{ruledtabular}
\begin{tabular}{cccccc}
$\Rsph/D$ & $\Lrod/D$ & $N$ & protocol & equil. sweeps & prod. sweeps \\
\hline
10 & 3  & 249  & density ramp & 3000 & 40000 \\
10 & 5  & 163  & density ramp & 3000 & 40000 \\
10 & 8  & 107  & density ramp & 3000 & 40000 \\
20 & 3  & 996  & density ramp & 3000 & 40000 \\
20 & 5  & 652  & density ramp & 3000 & 40000 \\
20 & 8  & 429  & density ramp & 3000 & 40000 \\
20 & 10 & 350  & density ramp & 3000 & 40000 \\
30 & 3  & 2241 & density ramp & 3000 & 40000 \\
30 & 5  & 1466 & density ramp & 3000 & 40000 \\
30 & 8  & 965  & density ramp & 3000 & 40000 \\
30 & 10 & 786  & density ramp & 3000 & 40000 \\
40 & 3  & 3984 & density ramp & 3000 & 40000 \\
40 & 5  & 2607 & density ramp & 3000 & 40000 \\
40 & 8  & 1716 & density ramp & 3000 & 40000 \\
40 & 10 & 1398 & density ramp & 3000 & 40000 \\
\end{tabular}
\end{ruledtabular}
\end{table}

All production runs sample density profiles every ten sweeps.

\subsection{Local smectic and chirality analysis}

For each rod $i$, the local smectic order parameter is
\begin{equation}
 \Psi_i(\dlay)=\frac{1}{|\mathcal N_i|}
 \sum_{j\in\mathcal N_i}
 \exp\!\left[
 2\pi i\,\frac{(\mathbf r_j-\mathbf r_i)\cdot\hat m_i}{\dlay}
 \right],
 \label{eq:si_psi}
\end{equation}
where $\mathcal N_i$ is the neighbour set inside the cutoff
$r_{\rm cut}$. The distances are ambient Euclidean chord distances in
$\mathbb R^3$, obtained from a fast spatial neighbour search (a
KD-tree, which partitions the rod-centre coordinates so that all
neighbours within $r_{\rm cut}$ are found without testing every pair),
and not geodesic arc distances measured along the spherical surface.
Because $r_{\rm cut}\ll\Rsph$, the chord and arc distances coincide to
leading order over a neighbourhood, so the chord metric is the natural
local choice. The trial layer normal
$\hat m_i$ is obtained by tilting the locked director in the local
tangent plane by an angle $\alpha$. The analysis maximises
$|\Psi_i|$ jointly over $40$ trial spacings and $29$ trial tilts in
the interval $[-35^\circ,35^\circ]$.

\begin{table}[h]
\caption{Neighbour cutoffs and trial spacing intervals used in the
local smectic-C analysis. For $\Lrod=3D$ and $5D$ the cutoff is
enlarged to $r_{\rm cut}=10D$ at the largest host $\Rsph=40D$.}
\label{tab:si_analysis}
\begin{ruledtabular}
\begin{tabular}{ccc}
$\Lrod/D$ & $r_{\rm cut}/D$ & trial $\dlay/D$ range \\
\hline
3  & 7  & $3.2$--$5.6$ \\
5  & 9  & $4.8$--$8.4$ \\
8  & 16 & $7.2$--$12.6$ \\
10 & 18 & $8.8$--$15.4$ \\
\end{tabular}
\end{ruledtabular}
\end{table}

The smectic-area fraction is the fraction of valid rods with
$|\Psi_i^\ast|>0.5$. The chirality index is
\begin{equation}
 \chi=\frac{|\langle\alpha\rangle|}{\sigma_\alpha},
 \label{eq:si_chi}
\end{equation}
where $\langle\alpha\rangle$ and $\sigma_\alpha$ are the mean and
standard deviation of the per-rod optimised tilts in the smectic patch.
The coherent-Sm-C classification uses $\chi\ge1.5$.

\end{document}